\documentclass[aps,prl,reprint,superscriptaddress]{revtex4-2}
\usepackage{xcolor}
\usepackage{graphicx}
\usepackage{bm}
\usepackage{braket}
\usepackage{amsmath}
\usepackage{svg}
\usepackage{siunitx}
\usepackage{relsize}
\usepackage{verbatim}

\usepackage{fancyhdr}
\begin{document}

\title{\textbf{Photonic qubit encoding interconversion for heterogeneous quantum networking} 
}%

\author{Vedansh Nehra}
\email{vedansh.nehra@techngs.com}
\affiliation{Technergetics LLC, Utica, NY 13502}

\author{Richard J. Birrittella}
\affiliation{Booz Allen Hamilton, Rome, NY 13441}

\author{Christopher C. Tison}
\affiliation{Air Force Research Laboratory, Rome, New York, 13441, USA}

\author{Benjamin K. Malia}
\affiliation{Technergetics LLC, Utica, NY 13502}

\author{Zachary S. Smith}
\affiliation{Air Force Research Laboratory, Rome, New York, 13441, USA}

\author{Dylan Heberle}
\affiliation{Technergetics LLC, Utica, NY 13502}

\author{Nicholas J. Barton}
\affiliation{Murray Associates of Utica, Utica, NY 13502}

\author{Amos Matthew Smith}
\affiliation{Air Force Research Laboratory, Rome, New York, 13441, USA}

\author{Andrew Brownell}
\affiliation{Murray Associates of Utica, Utica, NY 13502}

\author{Michael L. Fanto}
\affiliation{Air Force Research Laboratory, Rome, New York, 13441, USA}

\author{James Schneeloch}
\affiliation{Air Force Research Laboratory, Rome, New York, 13441, USA}

\author{Erin Sheridan}
\affiliation{Air Force Research Laboratory, Rome, New York, 13441, USA}

\author{David Hucul}
\affiliation{Air Force Research Laboratory, Rome, New York, 13441, USA}


\begin{abstract}
Quantum information processing, communication, and sensing networks are being developed with various qubit platforms that use different encoding schemes. Connecting quantum network nodes to distribute entanglement requires matching photon qubit basis encoding. In this work, we implement an interconversion protocol which converts photon qubit encoding from the polarization basis to the time-bin basis, transmits the photons through a transport fiber with large fluctuations in polarization, and converts back to polarization encoding for ease of measurement. This interconversion scheme faithfully transmits a polarization Bell state across the transport fiber by converting sources of infidelity to changes in transmission rate. These results illustrate a practical approach for interfacing distinct qubit platforms to enable modular and flexible operation in heterogeneous quantum networks.
\end{abstract}

\maketitle
\thispagestyle{fancy}



\begin{figure*}[t]
    \centering
    \includegraphics[width=\textwidth]{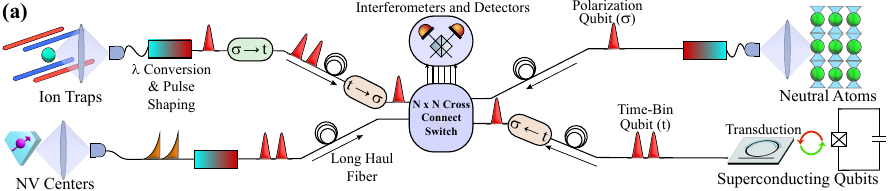}

    \vspace{1em}

    \includegraphics[width=\textwidth]{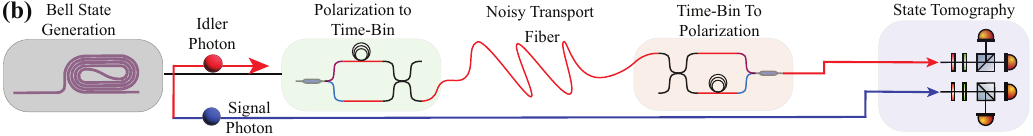}

    \caption{Representation of a heterogeneous quantum network and photonic qubit interconversion protocol. \textbf{(a)} Four example qubit platforms supporting varied photonic qubit basis encodings can interfere with each other via photon encoding interconversion. Additionally, encoding interconversion can be used to convert polarization encoding to time-bin encoding, suitable for long haul photon transmission. Time-bin qubits are converted back to polarization basis before detection for ease of measurement, as shown with photons from the ion trap node. \textbf{(b)} Idler photons from a polarization Bell state produced from a PIC source are sent through interconversion modules connected via a transport fiber. Signal photons are routed directly to a two qubit QST module for analysis.}
    \label{fig:overall_experiment}
\end{figure*}
Heterogeneous quantum networks (HQNs) with optical interconnects offer an avenue towards scaling quantum processing power\cite{Caleffi2024, Ang2024, monroe:2014, twofridge}, creating quantum sensor arrays \cite{Zhang2021, proctor2018multiparameter}, and connecting disparate quantum networks at a distance \cite{main2025distributed, jiang2007distributed, inns101km, delft_bell, covey:2023, inns_repeater, ion_REs}. To realize these applications, photons emitted by different qubit platforms must interfere to establish entanglement across network nodes. However, these photons may have different wavelengths, bandwidths, and encoding types. Even after matching wavelength and wavepacket shape, matching photon encoding type will be necessary.
 
For example, depending on the atomic species, trapped-ion platforms support several qubit encoding bases, including frequency, time-bin, and polarization degrees of freedom \cite{olmschenk2009quantum, saha2025high, blinov2004observation, matsukevich:2009, stephenson2020high, o2024fast}. On the other hand, the transduction of superconducting systems from the microwave to optical photons naturally makes use of time-bin encoding \cite{Kurpiers2019, Ilves2020, Sheridan2025b}, limiting the creation of high fidelity entanglement between superconducting and trapped ion qubits with different encoding types. Developing reliable qubit encoding interconversion protocols and modules as shown in Fig. \ref{fig:overall_experiment} \textbf{(a)} is therefore a key step towards linking otherwise incompatible quantum network nodes \cite{guccione2020connecting}. 

In addition, qubit basis conversion may aid in robust transmission of photonic qubits. Polarization-encoded photons are susceptible to time-dependent birefringence induced by environmental variations when transmitted over long distances in fibers, requiring active polarization stabilization to compensate for polarization drifts \cite{ulrich1979polarization, tan2024real, bersin2024, qunnectAPC}. Transferring to time-bin encoding (see photon emission from ion trap node in Fig. \ref{fig:overall_experiment} \textbf{(a)}) relaxes polarization control requirements \cite{valivarthi2014efficient,lago2023long,yu2020entanglement}.

In this work, we implement qubit basis interconversion modules for polarization and time-bin qubits at telecommunication wavelengths (see Fig.\ref{fig:overall_experiment} \textbf{(b)}). We generate polarization Bell states and show that by changing the qubit encoding from polarization to time-bin, we make the qubits less susceptible to polarization drifts. The time-bin encoded qubit is then transported over a short transport fiber wound around a polarization controller to induce birefringence. When converting back to polarization, variations in birefringence change the overall photon count rate rather than qubit state fidelity. This was confirmed by quantum state tomography (QST); measurements show the two-qubit Bell state fidelity remained approximately constant at 0.94 $\pm$ 0.01.

\begin{figure*}[t]
    \includegraphics[width=\textwidth]{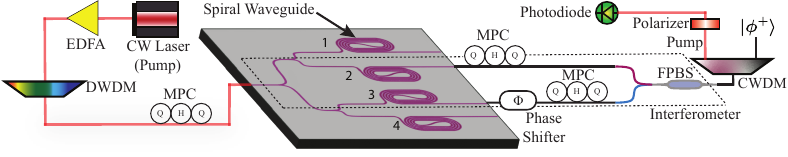}
    \caption{Diagram for polarization Bell state generation using PIC. Amplified pump light is coupled to the chip and distributed via cascaded $1 \times 2$ splitters. Spectrally entangled biphotons generated by SFWM from two spiral waveguides, having the state $\ket{H_sH_i}$, are combined using an FPBS. The relative phase between the spirals is stabilized to generate the Bell state $\ket{\phi^+}$ (Eq. \ref{eq:bell_state_spiral}). To increase Bell state fidelity, we balance attenuation between both arms of the interferometer by controlling the amount of light that gets coupled to the output of the FPBS using manual polarization controllers (MPC).}
    \label{fig:bell_gen}
\end{figure*}

We use a photonic integrated circuit (PIC) to generate pairs of entangled photons (biphotons) via spontaneous four-wave mixing (SFWM) in silicon waveguides (see Fig. \ref{fig:bell_gen}) \cite{takesue2007entanglement}. Two pump photons from a narrow-band, continuous wave (CW) laser at \SI{1550}{\nm} produce phase-matched signal-idler photon pairs of the same polarization with spectrally anti-correlated frequencies $\omega_s$, $\omega_i$. We spectrally filter the signal and idler photons around \SI{1530}{\nm} and \SI{1570}{\nm}, respectively.

The PIC utilizes cascaded $1 \times 2$ splitters \cite{olislager2013silicon, silverstone2014chip} with four outputs \cite{Sheridan2025a}. We use two outputs from the PIC to generate a polarization entangled Bell state using an interferometric setup shown in Fig. \ref{fig:bell_gen}. A CW pump laser is routed through an erbium-doped fiber amplifier (EDFA) and spectrally filtered using a dense wavelength division multiplexer (DWDM) (ITU standard C34 channel at 1550.12 nm). Since the biphoton production efficiency from one spiral waveguide is low ($\eta = 10^{-12}$ biphotons per pump pair), the probability of simultaneously generating biphotons in both spirals 2 and 3 in Fig. \ref{fig:bell_gen} is of order $\epsilon = 10^{-24}$. Thus, pumping both spirals 2 and 3 via the cascaded $1 \times 2$ splitters leads to a path-entangled N00N state \cite{silverstone2014chip} described by:

\begin{equation}
    \ket{N00N} = \frac{1}{\sqrt{2}}\Big(\ket{\psi,0}_{2,3} + \ket{0,\psi}_{3,2}\Big)
\end{equation}

\noindent where $\ket{\psi} = \ket{H_sH_i}$ and $\ket{\psi,0}_{2,3}$ ($\ket{0,\psi}_{3,2}$) indicates biphoton generation in spiral 2 (3) and no biphoton generation in spiral 3 (2). A fiber polarizing beam splitter (FPBS) combines photons from the two spirals, with manual polarization controllers (MPCs) in a Quarter-Half-Quarter waveplate (QHQ) configuration before the FPBS inputs to rotate polarization and control photon transmission through the FPBS, thereby balancing $H_sH_i$ and $V_sV_i$ photon counts. The resulting  output is a polarization entangled Bell state:
\begin{equation}
    \ket{\phi^+} = \frac{1}{\sqrt{2}}(\ket{H_{s}H_{i}} + e^{i\phi}\ket{V_{s}V_{i}})
    \label{eq:bell_state_spiral}
\end{equation}
\noindent where $\phi$ denotes the relative phase between the biphotons generated from the two spirals and $H_s$ ($V_s$), $H_i$ ($V_i$), denotes the horizontal (vertical) polarization of the signal and idler photons respectively.

The FPBS output is routed through a coarse wavelength division multiplexer (CWDM) with channel bandwidths of \SI{8}{\nm}, center wavelength separations of \SI{20}{\nm}, adjacent channel isolation $>\SI{30}{\dB}$ and non-adjacent channel isolation of $>\SI{40}{\dB}$. We use the $1530\;\mathrm{nm}$ and $1570\;\mathrm{nm}$ channels to select the signal (s) and idler (i) wavelengths at \SI{1530}{\nm} and \SI{1570}{\nm}, respectively, and the CWDM directs the \SI{1550}{\nm} pump light to a photodiode through a polarizer. The resulting interference of the pump light provides feedback to the phase shifter to stabilize relative path length difference of the interferometer arms \cite{yu2020entanglement}, thereby stabilizing $\phi$. The idler photons pass through a $1\;\mathrm{nm}$ bandwidth filter set at \SI{1570}{\nm}, and the signal photons go through two \SI{1530}{\nm} CWDM channels to further filter the pump from the signal-idler outputs.

To quantify quality of entanglement, we perform a Bell's inequality measurement \cite{aspect1981experimental} by taking coincidences between signal and idler photons using a Superconducting Nanowire Single Photon Detector (SNSPD) (jitter $\sim$ \SI{20}{\pico\second}) and a time-tagger (\SI{1}{\pico\second} resolution). After subtracting accidental coincidences, we obtain a Clauser–Horne–Shimony–Holt (CHSH) parameter $S = 2.79\;\pm\; 0.01$ and  a fidelity of $0.996 \pm 0.001$ as shown in Fig. \ref{fig:combined}.

\begin{figure*}[t]
    \includegraphics[width=\textwidth]{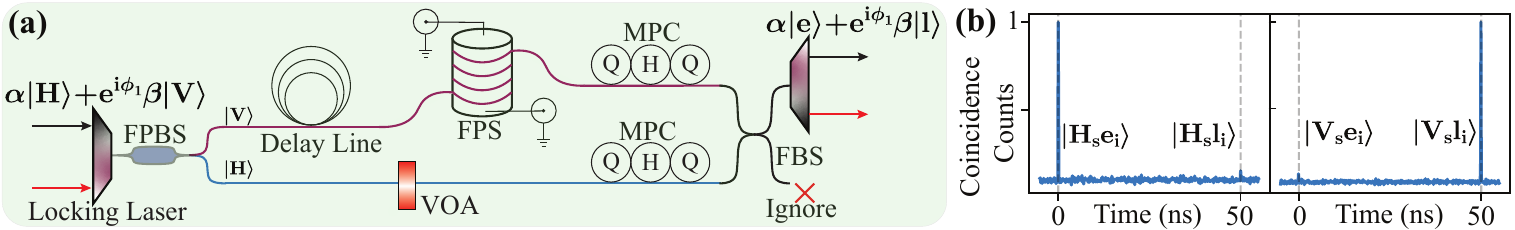}
    \caption{Schematic for polarization to time-bin conversion and photon coincidence results. \textbf{(a)} An FPBS separates $\ket{H}$ and $\ket{V}$ polarized photons. $\ket{V}$ polarized photons are sent through a \SI{10.2}{\meter} delay line to add extra path length relative to $\ket{H}$ photons, thereby creating time-bins separated by \SI{50}{\ns}. The two paths are recombined using an FBS, with MPCs in both arms to make the polarizations of both the time-bins the same. Active control to the FPS stabilizes the interferometric phase. A VOA equalizes attenuation across both arms, minimizing polarization dependent loss. \textbf{(b)} Early ($\ket{e}$) and late ($\ket{l}$) time-bins separated by \SI{50}{\ns} observed by taking coincidences of signal and idler H (early) and V (late) photons.}
    \label{fig:pol-tb}
\end{figure*}

Having demonstrated a high-quality source of polarization entangled Bell states, we now focus on interconverting this state between polarization and time-bin encoding for quantum networking. As shown in Fig. \ref{fig:overall_experiment} \textbf{(b)}, the idler photons are sent through through a polarization to time-bin interconversion module followed by a time-bin to polarization interconversion module, connected by \SI{4}{\meter} of transport fiber. To simulate the polarization drift often observed in long-distance quantum networks \cite{Sheridan2025a, bersin2024}, three manual polarization paddles (MPPs) in a QHQ configuration strain this fiber. Interconversion modules are enclosed to minimize air currents.

An asymmetric Mach-Zehnder interferometer (AMZI) converts the polarization qubit to a time-bin qubit as shown in Fig. \ref{fig:pol-tb} \textbf{(a)}. An FPBS separates orthogonal polarizations into different paths, where one path is \SI{10.2}{\meter} longer than the other to create two time-bins separated by \SI{50}{\nano\second}. In order to demonstrate compatibility of these modules with photons generated by atom-based qubits with temporal widths on the order of tens of nanoseconds, a relatively large time-bin separation of \SI{50}{\nano\second} was chosen such that the time-bins have low temporal overlap as shown in Fig. \ref{fig:pol-tb} \textbf{(b)}.

The two AMZI paths are recombined on a $2 \times 2$ fiber beam splitter (FBS), with the transport fiber connected to one of the output ports. MPCs are also used in both arms of the AMZI to ensure that both time-bins have the same polarization, making all degrees of freedom indistinguishable except time of arrival. A variable optical attenuator (VOA) is used in one arm to match attenuation in both arms to minimize polarization dependent loss (PDL). For a polarization qubit $\ket{\psi_{Pol}} = \alpha \ket{H} + e^{i\phi}\beta\ket{V}$, the conversion module transforms $\ket{\psi_{Pol}}$ into a time-bin qubit $\ket{\psi_{TB}} = \alpha \ket{e} + e^{i\phi_1}\beta\ket{l}$, where $\phi_1 = \phi + \phi'$ and $\phi'$ is the acquired phase difference between the $\ket{e}$ (early) and $\ket{l}$ (late) photons as they travel through their respective arms. The converted Bell state in Eq. \ref{eq:bell_state_spiral} after the idler photon passes through the AMZI is:
\begin{equation}
    \ket{\psi_{Conv}} = \frac{1}{\sqrt{2}} (\ket{H_s e_i} + e^{i\phi_1}\ket{V_s l_i}),
\end{equation}
and the separation between the time-bins is shown in Fig. \ref{fig:pol-tb} \textbf{(b)}.

Since the two orthogonal polarizations represent the basis states of the input qubit, any drift in the differential path length directly maps onto a change in this relative phase, which in turn alters the time-bin encoded qubit state. We stabilize this AMZI using a \SI{1367}{\nm} DFB laser with an FWHM less than \SI{10}{\MHz} locked to the  $5P_{3/2}\;,F = 4 \rightarrow \; 6S_{1/2}\;,F' = 3 $ transition in $^{85}\text{Rb}$. The locking laser is multiplexed into and out of the interferometer using the expansion port of a CWDM shown in Fig. \ref{fig:pol-tb} \textbf{(a)}. After the FBS the photodiode signal is fed back to a fiber piezo stretcher (FPS). To estimate the phase stability, Stokes parameters of the stabilization laser (after passing through the AMZI) are measured using a polarimeter and used to calculate the phase difference between orthogonal polarization components using Jones vectors \cite{jones1941new}. The standard deviation of phase is observed to be \SI{1}{\degree} over \SI{5}{\min} at \SI{1367}{\nm}, implying $\phi_1$ has a standard deviation less than \SI{1}{\degree} at \SI{1570}{\nm}. As shown in Fig. \ref{fig:pol-tb} \textbf{(b)}, upon detecting a H (V) polarized signal photon and a subsequent idler photon, the idler photon is an early (late) time-bin with probability $94 \pm 1 \%$ ($96 \pm 1 \%$). Imperfect interconversion can be attributed to non-zero PDL due to manually setting MPCs as well as extinction ratios of FPBS and FBS. 

\begin{figure*}
    \includegraphics[width=\textwidth]{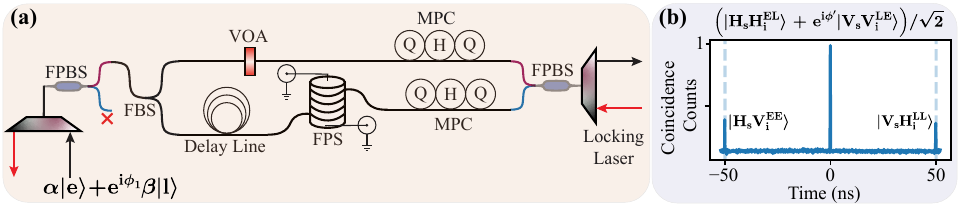}
    \caption{Schematic for time-bin to polarization conversion and photon coincidence results. \textbf{(a)} The FPBS allows a fixed polarization through to the interconversion module. An FBS separates photons into two paths having a relative path length of \SI{10.2}{\meter}, which are combined using an FPBS to convert back to polarization basis, with MPCs in both arms ensuring maximum transmission through the final FPBS. \textbf{(b)} Plot showing the three arrival times of the time-bin encoded idler photons that pass through a transport fiber and then the time-bin to polarization interconversion module. Using an FBS distributes both $\ket{e}$ and $\ket{l}$ photons to the two arms of the AMZI. By post-selecting photons traversing a combination of short and long paths (indicated by the middle peak), we complete the time-bin to polarization conversion and retrieve the polarization qubit.}
    \label{fig:tb-pol}
\end{figure*}

After converting the polarization qubit to a time-bin qubit, the idler photon passes through the transport fiber and enters the time-bin to polarization conversion module. Since both the early and late time-bins have the same polarization, and the timescale of polarization drifts of the transport fiber is much larger than the temporal separation between the time-bins (\SI{50}{\ns}), we assume both time-bins are affected by polarization variations in the same manner  \cite{Sheridan2025a, bersin2024}. We use an FPBS at the beginning of the time-bin to polarization interconversion module to ensure transmission of a fixed polarization of light into the AMZI as shown in Fig. \ref{fig:tb-pol} \textbf{(a)}. The AMZI begins with a $1 \times 2$ beam splitter with the input port connected to one of the outputs of the FPBS. The two arms once again have \SI{10.2}{\m} of relative path length difference, with one of the arms having an FPS for phase stabilization. MPCs are used before the FPBS to ensure maximum transmission of photons to the output port.

After the photons are converted back to polarization encoding, there are three possible arrival times of the photons \cite{ono2024quantum,pilnyak2019quantum} as shown in Fig. \ref{fig:tb-pol} \textbf{(b)}. These arise from the combination of short and long paths in the two AMZIs. The earliest arrival corresponds to the photon taking short paths in both interferometers (early-early or EE), while the latest arrival indicates that the photon took the long path in both the modules (late-late or LL). However, if the photon travels through the short path in the first interferometer and through the long path in the second interferometer (or vice-versa), its time of arrival will be exactly in between EE and LL, meaning the photon has taken opposite paths in both the modules. Hence, by post-selecting this intermediate time, the polarization qubit can be faithfully recovered, as shown in Fig. \ref{fig:tb-pol} \textbf{(b)}, where the middle peak corresponds to the correct polarization state given by:
\begin{equation}
    \ket{\psi} = \frac{1}{\sqrt2}\Big(\ket{H_sH_i^{{EL}}} + e^{i\phi'}\ket{V_sV_i^{{LE}}}\Big),
    \label{eq:end_state}
\end{equation}
where $\ket{H_sH_i^{EL}}$ ($\ket{V_sV_i^{LE}}$) indicates the H (V) idler photon taking the short (long) path followed by the long (short) path. 

Accurate temporal matching is required to recover maximum interference at the QST tomography module. Group delays in the AMZIs must be set so that the early and late components overlap within the temporal width of the photon when they are recombined; this is most evident in the diagonal/anti-diagonal (A/D) polarization basis projections, whose visibility is directly tied to the temporal overlap of the two paths. To ease matching the temporal overlap of the time-bins for interference, we apply a narrowband filter with \SI{1}{\nm} bandwidth to the idler photon with a bandwidth of (\SI{20}{\mm}) set by the CWDM. This lengthens the photon wavepacket from \SI{0.6}{\ps} (\SI{0.18}{\mm}) to \SI{12.3}{\ps} (\SI{3.7}{\mm}), enabling higher visibility interference (at the expense of loss) in this experiment. To ensure maximal temporal overlap of the late-early and early-late arrival times, we use a digitally controlled fiber delay line (not shown in \ref{fig:tb-pol} \textbf{(a)}) in the shorter arm of the second AMZI, which allows us to tune temporal mismatches up to \SI{660}{\ps} with \SI{1}{\ps} resolution. 

To characterize the entire interconversion setup, where polarization qubits are converted to time-bin qubits and back to polarization qubits, we perform two-qubit QST while continuously varying the strain of the transport fiber to induce polarization rotation. Two-qubit QST is conducted by sending only the idler photons through both interconversion modules while coincidences are taken between the signal and idler photons. A single overcomplete tomography measurement is performed by taking 36 total measurement results \cite{altepeter20044} spanning the canonical bases (H/V, D/A and R/L basis). A Linear Inversion algorithm is used for reconstructing the density matrix \cite{WiedQuantumST} and calculating fidelity of the Bell state. To estimate uncertainties in the reconstructed density matrices and fidelity, we use a Monte Carlo procedure \cite{altepeter20044}.

\begin{figure}[t]
    \includegraphics[width=\columnwidth]{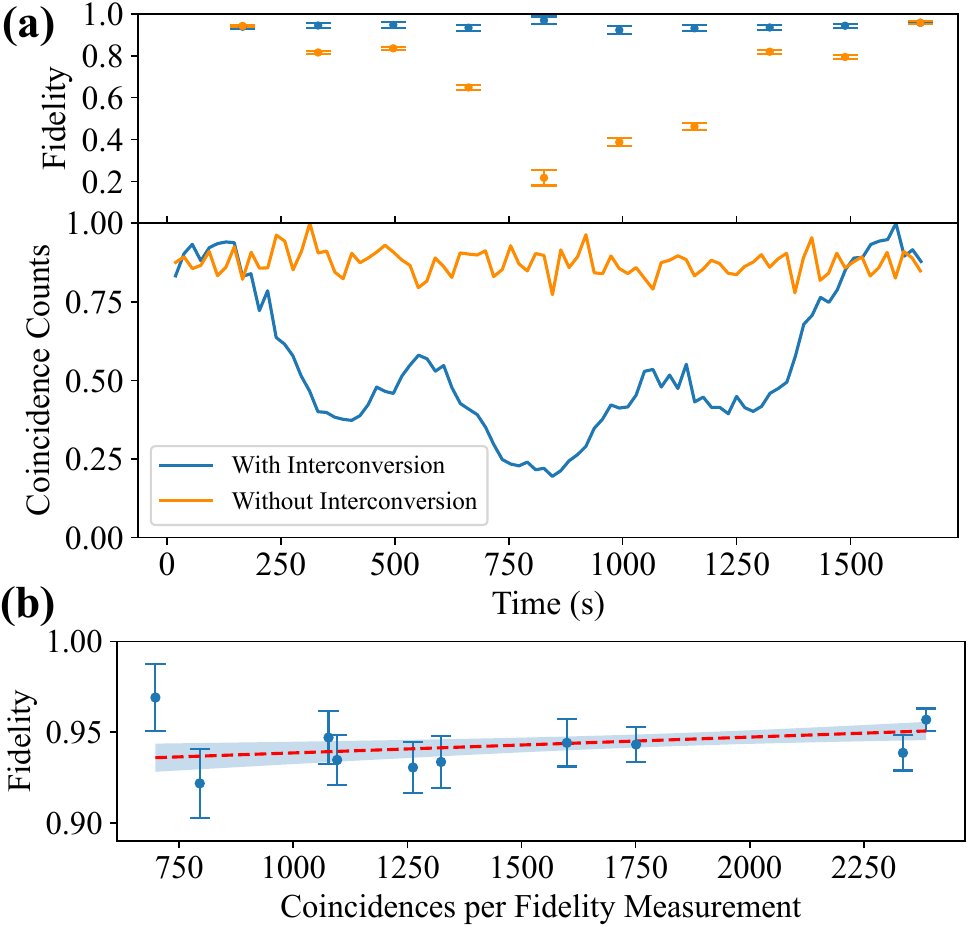}
    \caption{Fidelity and transmission rate variation for time-bin and polarization encoded photons. \textbf{(a)} Coincidence counts (normalized) of signal and idler photons as polarization in the transport fiber is perturbed by rotating the HWP in the MPP by \SI{1}{\degree} every \SI{5}{\s}. Idler photons passing through the interconversion modules (connected via transport fiber) show huge changes in the transmission rate while the fidelity remains fairly constant (0.94 average fidelity), whereas idler photons passing through only the transport fiber show reduction in fidelity as the polarization changes while the coincidence rate remains fairly constant. \textbf{(b)} Each fidelity measurement is plotted against total coincidences, and a linear fit gives a slope of $8.7\times 10^{-6}$, signifying minimal dependence of fidelity on photon transmission rate.}
    \label{fig:final result}
\end{figure}

While the idler photons pass through the transport fiber (see Fig. 1(b)), we strain the optical fiber using MPPs. The strain, caused by rotating the MPP half-waveplate serves as a polarization perturbation in between the two interconversion modules, resulting in a change in the measured coincidence counts (due the FPBS at the beginning of the time-bin to polarization interconversion), as shown in Fig. \ref{fig:final result} \textbf{(a)}. While performing tomography, the half-waveplate is rotated step-wise by \SI{1}{\degree} every \SI{5}{\s}. The waveplate starts at \SI{0}{\degree}, and upon reaching an angle of \SI{160}{\degree}, rotates back to \SI{0}{\degree}.  

As shown in Fig. \ref{fig:final result} \textbf{(b)}, the average fidelity of the final polarization Bell state (Eq. \ref{eq:end_state}) across ten tomography measurements spanning $\approx \SI{26}{\min}$ is $0.94 \pm 0.01$ while the coincidence rate varies due to varying strain in the transport fiber (Fig. \ref{fig:final result} \textbf{(a)}). This is compared with the case when idler photons are kept in the polarization basis and sent through the transport fiber and rotated in the same manner as described above. In this scenario, the total number of coincidence counts remains approximately constant (Fig \ref{fig:final result} \textbf{(a)}), whereas fidelity of the measured Bell state changes, indicating that fluctuations in rate can be traded for stability in fidelity.

We attribute the variation of the final Bell state fidelity primarily due to drifts in the interferometer phase ($\approx 2\%$). In addition, we attribute the decrease in the final Bell state fidelity  (from $\mathcal{F} =  0.996$) to drifts in the interferometer phase ($\approx 1\%)$, extinction ratio of FPBS ($\approx 1\%$) and imperfectly setting the MPCs, VOAs and SNSPD detector efficiencies ($\approx 8\%$) to balance PDL. These errors are consistent with the observed decrease in overall fidelity. At the cost of complexity, loss in interconversion modules in this work (\SI{\approx12}{\dB} and \SI{\approx7}{\dB}) can be decreased by up to \SI{6}{\dB} by using active switches (e.g. electro-optic modulators) instead of beamsplitters to combine/split time-bins.

In conclusion, we have shown that by interconverting between polarization and time-bin encoding, we enable robust transmission of photonic qubits with minimal decrease in fidelity over a polarization-distorting link without requiring active feedback over the transmission fiber. 

Additionally, interconversion between encoding schemes can allow distinct quantum hardware platforms which emit photonic qubits with otherwise incompatible bases to interface with each other, paving a way towards heterogeneous quantum networks. For example, superconducting systems emitting transduced photons encoded in time-bins could be converted to the polarization basis to interface with polarization encoded photons emitted from atomic ensembles, ions, or NV centers \cite{Sheridan2025b, sun2022deterministic, tchebotareva2019entanglement, jayakumar2014time}, integrating systems with complementary capabilities within the same network infrastructure.

This technology was primarily supported by the Microelectronics Commons Program, a DoW initiative, under award number N00164-23-9-G061. We acknowledge the contributions of Stefan Preble for designing the PIC spiral source chips. The views and conclusions contained herein are those of the authors and should not be interpreted as necessarily representing the official policies or endorsements, either expressed or implied, of the United States Air Force, the Air Force Research Laboratory or the U.S. Government. This document is approved for public release, distribution unlimited. PA \#AFRL-2026-1512.

\bibliography{library}
\newpage

\appendix

\section{Appendix: Bell State Generation}
\begin{figure}[b]
    \includegraphics[width=\columnwidth]{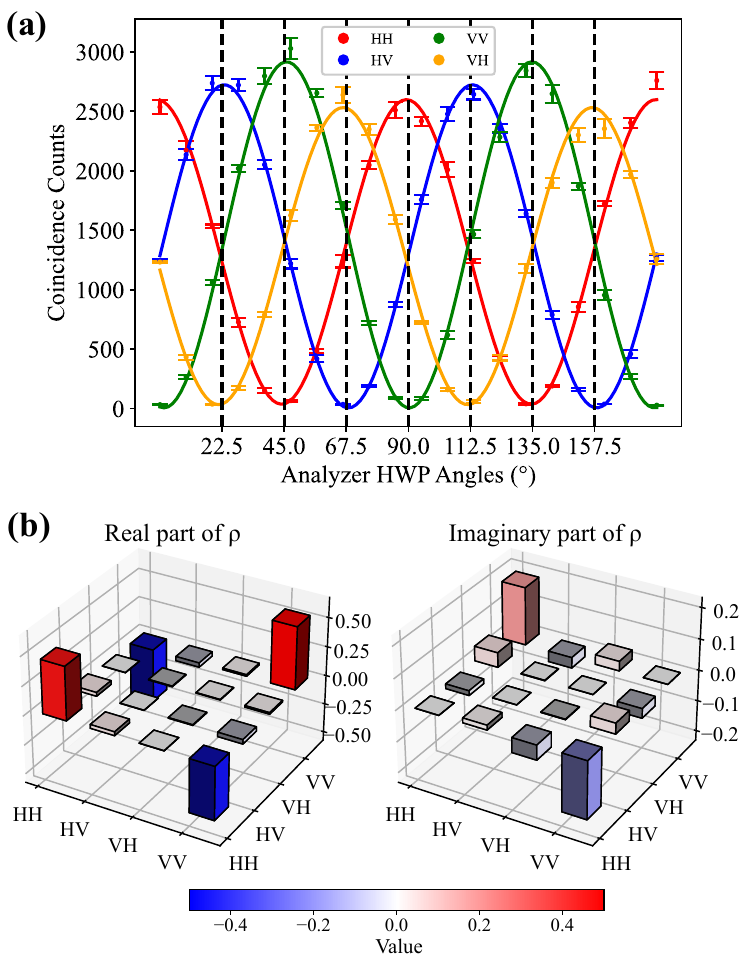}
    \caption{\textbf{(a)} CHSH violation (after subtracting accidental coincidences) and \textbf{(b)} Two qubit state tomography results of the Bell state directly from the PIC source. Coincidences are taken between signal and idler photons and a linear inversion method algorithm is used for density matrix reconstruction \cite{WiedQuantumST} which is used to calculate fidelity with the Bell state given in Eq. \ref{eq:bell_state_spiral}. (HWP: Half-Wave Plate)}
    \label{fig:combined}
\end{figure}

We use the Quantum State Tomography module to perform a Bell's inequality test to quantify the quality of entanglement for the state:
\begin{equation}
    \ket{\phi^+} = \frac{1}{\sqrt{2}}(\ket{H_{s}H_{i}} + e^{i\phi}\ket{V_{s}V_{i}})
\end{equation}

The tomography module's quarter-waveplates are kept at \SI{0}{\degree}. The half-waveplates are rotated to specific angles and coincidence counts are recorded at various combinations of angles, as depicted by the dashed lines in Fig. \ref{fig:combined}, to measure correlations. Error bars represent the standard deviation of five repeated coincidence measurements.

An overcomplete tomography measurement is performed by taking 36 total measurement results \cite{altepeter20044} spanning the canonical bases (H/V, D/A and R/L basis). A Linear Inversion algorithm is used for reconstructing the density matrix \cite{WiedQuantumST} and calculating fidelity of the Bell state. To estimate uncertainties in the reconstructed density matrices and fidelity, we use a Monte Carlo procedure \cite{altepeter20044}.

\end{document}